\documentstyle[12pt,epsf,epsfig]{article}
\def\gappeq{\mathrel{\rlap {\raise.5ex\hbox{$>$}}
{\lower.5ex\hbox{$\sim$}}}}
\def\permil{$\%\raise.20ex\hbox{$_0$}}
\def\lappeq{\mathrel{\rlap{\raise.5ex\hbox{$<$}}
{\lower.5ex\hbox{$\sim$}}}}
\begin{document}
\topmargin -1.0cm
\oddsidemargin -0.8cm
\evensidemargin -0.8cm
\pagestyle{empty}
\begin{flushright}
CERN-TH/97-129
\end{flushright}
\vspace*{10mm}
\begin{center}
{\Large\bf Supersymmetry Breaking with Vanishing $F$-terms}\\
\vspace{0.5cm}
{\Large\bf  in Supergravity Theories}\\
\vspace{2cm}
{\large\bf Gia Dvali$^{(1)}$  and Alex Pomarol$^{(2)}$}\\
\vspace{.7cm}
$^{(1)}${Theory Division, CERN}\\
{CH-1211 Geneva 23, Switzerland}\\
\vspace{0.3cm}
$^{(2)}${Institut de F\'\i sica d'Altes Energies}\\
{Universitat Aut\`onoma de Barcelona}\\
{E-08193 Bellaterra, Barcelona, Spain}\\
\end{center}
\vspace{2cm}
\begin{abstract}
In conventional supergravity theories, supersymmetry is broken by
a non-zero $F$-term, and the cosmological constant is fine tuned to zero
by a constant in the superpotential $W$. 
We discuss a class of supergravity theories with vanishing $F$-terms
but $\langle W\rangle \not= 0$ being generated dynamically.
The cosmological constant is assumed to be cancelled by a non-zero
$D$-term. In this scenario the gravity-mediated soft masses depend only on
a single parameter, the gravitino mass. They are automatically
universal, independently of the K\"ahler metric, and real.
Thus, dangerous flavor or CP violating interactions are suppressed.
Unlike in conventional supergravity models, the Polonyi problem
does not arise.
\end{abstract}

\vfill
\begin{flushleft}
CERN-TH/97-129\\
June 1997
\end{flushleft}
\eject
\pagestyle{empty}
\setcounter{page}{1}
\setcounter{footnote}{0}
\pagestyle{plain}
 

\section{Introduction}
One of the major challenges in supersymmetric models for 
particle physics is the understanding of the
breaking of supersymmetry.
Supergravity theories offer 
one of the  simplest  scenarios \cite{sugra}.
Supersymmetry is broken by an $F$-term of a hidden
superfield, and gravity plays the role of transmitting the
supersymmetry breaking to the ordinary quarks and leptons.
Nevertheless,
the induced soft breaking masses 
are not  generically universal \cite{soni} nor
real \cite{edm}, 
leading to dangerous  flavor and CP violating interactions. 

Here we will  consider a different scenario of 
supergravity breaking.
The $F$-terms of the hidden sector superfields will be zero, but not their
superpotential. This will be generated dynamically at a low-energy  scale.
By setting the   cosmological constant to zero (with a 
non-zero $D$-term), 
the induced superpotential  
will parametrize the   breaking of supersymmetry.
Like in conventional supergravity models, gravity will mediate
the breaking of supersymmetry  to the ordinary
quarks and leptons.
Nevertheless, 
in our scenario
the scalar masses turn to be automatically universal, independently of 
the K\"ahler metric.
The origin of  universality is  a
consequence of the super-Weyl-K\"ahler 
symmetry  of supergravity.
We find that the 
trilinears and  gaugino masses are zero at tree-level.
Nevertheless, gaugino masses 
 are generated at the one-loop level if heavy states are present
in the theory. 
This is the case of grand unify theories (GUTs) where
 the large number of heavy states can induce sizeable gaugino
masses. 
We will show that our scenario do not suffer 
from the supersymmetric CP problem,
since no extra
CP-violating phases are generated.

\section{Dynamical Supersymmetry Breaking 
with $\langle F\rangle = 0$ and 
$\langle W\rangle\not= 0$}

In supergravity theories the scalar potential is given by \cite{sugrapo}
\begin{equation}
V=K_{i\bar\jmath}F^iF^{\dagger\bar\jmath}
-3e^{K/M^2_P}\frac{|W|^2}{M^2_P}
+\frac{1}{2}D_aD_a\, ,
\label{potential}
\end{equation}
where $K$ is the K\"ahler potential, $K_{i\bar\jmath}$ is the K\"ahler 
metric, $W$ is the superpotential,
\begin{equation}
F^{\dagger\bar\jmath}=-e^{K/2M^2_P}K^{i\bar\jmath}
\left(\partial_iW+\partial_iK\frac{ W}{M^2_P}\right)\, ,
\label{fterm}
\end{equation}
are the generalized $F$-terms of the 
chiral superfields
and $D_a$ are the gauge $D$-terms of the vector superfields.
We will consider models in which
 $\langle F_i\rangle=0$  
but $\langle W\rangle\not=0\ll M^3_P$.
The negative contribution to the 
 cosmological constant coming from the second term of eq.~(\ref{potential})
will be assumed to be cancelled out 
 by a non-zero  $D$-term. 
We will not specify how 
such a $D$-term is generated. We will rather take this
 fine tuning as an assumption.
We will assume that the light states
are neutral under this $D$-term, and therefore that
this $D$-term does not play
any role in the breaking of supersymmetry 
in the ordinary fields.
In these models, it will be
the  non-zero gravitino mass $m_{3/2}=\langle e^{K/2M^2_P}
W\rangle/M^2_P$  the source of low-energy supersymmetry breaking.

A general 
(sufficient) condition for generating
$\langle W\rangle \neq 0$ and  vanishing $F$-terms  can be 
easily formulated 
in supergravity theories:
Any model of global supersymmetry with
 chiral superfield(s)
$X$ that has a minimum that satisfies

(a) $\langle X\rangle = X_{global}\ll M_P$ and   non-zero
 $ \langle W(X_{global}) \rangle\ll M^3_P$,

(b) preserves global supersymmetry, $\langle \partial_XW\rangle=0$, and

(c) do not have
 chiral superfields 
with  masses smaller than ${\langle W \rangle / M_P^2}$,

\noindent will have,
in the locally supersymmetric extension,
a local minimum at some $\langle X\rangle = X_{grav}$
with vanishing generalized $F$-terms and 
$ \langle W(X_{grav}) \rangle\not=0$.

The proof goes as follows. 
Due to (a) and (c) the gravity 
corrections to the global supersymmetric
masses and VEVs
are small.
Therefore, these corrections cannot create a massless state, the would-be 
Goldstino, and supergravity cannot be broken, {\it i.e.}, 
the  $F$-terms must vanish.

Examples of dynamically generated superpotentials that do not break
 supersymmetry in the flat limit can be readily constructed. 
Models of this type are any strongly coupled
SU(N) theory with a  number 
of flavors less than N that are massive \cite{nonpert}.
The simplest is SU(2) with one flavour
(two doublets)  $\psi, \bar{\psi}$. 
This theory has an unique $D$-flat 
direction that can be parametrized by a single holomorphic invariant 
(``meson'') $X^2 = \bar{\psi}\psi$. Far along the flat direction,
the SU(2) is completely
broken and the only low-energy (super)field is  the
massless $X$. 
By adding a mass term in the 
superpotential
\begin{equation}
W = mX^2\, ,
\end{equation}
one can lift all the points of the vacuum
but the origin ($X=0$) in which the gauge symmetry is unbroken. 
Instantons, however, 
generate a superpotential which pushes 
$X$ away from the origin \cite{ads}:
\begin{equation}
W = mX^2 + \Lambda^5 / X^2\, ,
\end{equation}
where $\Lambda$ is the strong scale of the SU(2).
This stabilizes $X$ at
\begin{equation}
\langle X^2\rangle = \sqrt{\Lambda \over m}\Lambda^2\, ,
\end{equation}
and generates a non-zero 
vacuum expectation value (VEV) for the superpotential 
\begin{equation}
\langle W\rangle = 2\sqrt{\Lambda m}\Lambda^2\, ,
\end{equation}
without breaking  supersymmetry. For $m\ll \Lambda$,
$X$ can be treated as
a canonically normalized superfield with mass $4m$. Therefore,
there are not
 massless states in the theory,
 and the conditions (a), (b) and  (c) hold.
 It can be verified by explicit minimization that 
$\langle F_X\rangle$ is also zero
 in the supergravity case, 
in agreement with the general argument given above.
By continuity this is also true for  $m\gappeq\Lambda$.

Another examples 
can be found in superpotentials generated  from
a quantum modified moduli space \cite{nonpert}, {\it e.g.} \cite{savas}:
\begin{equation}
  \label{quantum}
W=\lambda X^3+\Lambda^2 X \, , 
\end{equation}
where $\Lambda$ is the meson condensate of a strong SU(N) with a 
number of flavors equal to N.
Also in these models we have vanishing $F$-terms and 
$\langle W\rangle\not=0\ll M^3_P$. 

\section{Soft Supersymmetry Breaking Terms}

\subsection{Gravity-induced soft terms}

In supergravity theories the scalar soft masses are given by
(for canonically normalized  fields) \cite{soni}
\begin{equation}
m^2_{i\bar\jmath}=-2m^2_{3/2}\delta_{i\bar\jmath}+
\frac{\langle F^l\rangle \langle F^{\dagger\bar k}\rangle}
{M^2_P}\left[K_{l\bar k}\delta_{i\bar\jmath}-
R_{l\bar ki\bar\jmath}\right]\, ,
\label{masses}
\end{equation}
where $R_{l\bar ki\bar\jmath}$ is the curvature in  the K\"ahler manifold.
The last term in eq.~(\ref{masses}) is responsible for the breaking
of universality and inducing flavor violating interactions
in the squark sector. 
One popular way 
 to  retain universality is to assume that   the K\"ahler 
manifold is flat (minimal supergravity) \cite{sugra}. Nevertheless, 
 there is  a priori no reason, based on symmetry principles, 
to expect that (from string theories one learns that 
universality can only arise under certain conditions \cite{string}).

In theories with vanishing $\langle F_i\rangle$, 
universality is automatic
since the last term of eq.~(\ref{masses}) is zero. 
In these theories, we obtain
\begin{equation}
\label{soft}
m^2_{i\bar\jmath}=-2m^2_{3/2}\delta_{i\bar\jmath}\, .   
\end{equation}
The origin of  universality can be understood as a 
consequence of the super-Weyl-K\"ahler 
symmetry 
\begin{eqnarray}
W&\rightarrow & e^{-{\cal F}/M^2_P}W\, ,\nonumber\\
K&\rightarrow & K+{\cal F}+{\cal F}^\dagger\, ,
\end{eqnarray}
where ${\cal F}$ is an arbitrary 
holomorphic function of chiral superfields.
To see that, let us  write the effective supergravity Lagrangian 
in a flat gravitational background,
using the superfield formalism \cite{compensator, anomaly, bagger}:
\begin{equation}
  \label{supergravity}
{\cal L}=-3M^2_P\int d^4\theta \varphi\varphi^\dagger e^{-K/3M^2_P}+
\left(\int d^2\theta  \varphi^3W+h.c.\right)\, ,
\end{equation}
where  $\varphi$ is the compensator superfield; this is an auxiliary
(non-dynamical) chiral superfield
introduced in the theory to
make  the super-Weyl-K\"ahler  symmetry manifest. For this purpose,
$\varphi$ should transform
 as 
\begin{equation}
\varphi\rightarrow  e^{{\cal F}/3M^2_P}\varphi\, ,
\end{equation}
under
the super-Weyl-K\"ahler  transformation.
The compensator superfield
 can be eliminated by its equation of motion that gives \cite{bagger}
\begin{equation}
  \label{ftermphi}
\varphi=e^{K/6M^2_P}\left[1+\theta^2\left(e^{K/2M^2_P}\frac{W^\dagger}
{M^2_P}+\frac{\partial_iK\, F^{i}}{3M^2_P}\right)\right]\, .
\end{equation}
{}For
$\langle W\rangle\not=0$,  $\varphi$ breaks supersymmetry.
The super-Weyl-K\"ahler symmetry  guarantee that 
$\varphi$ couples universally to matter and therefore that
the scalar fields receive universal
soft masses.
The result of eq.~(\ref{soft}) can be easily obtained by inserting
eqs.~(\ref{fterm}) and (\ref{ftermphi}) 
in  eq.~(\ref{supergravity}).

Trilinears and  bilinears soft terms can be also obtained by 
the same procedure. One obtains 
\begin{eqnarray}
&&A_{ijk}=0\, ,\label{trilinear}\\
&&B_{ij}=-m_{3/2}M_{ij}\, ,\label{bilinear}
\end{eqnarray}
where $M_{ij}$ are the bilinear couplings in the superpotential.
Since the trilinears vanish, 
there are not
 left-right flavor-violating interactions\footnote{
It is interesting to mention that these
interactions are usually present even in theories with
non-Abelian
flavor symmetries and can lead to large dipole moment transitions
\cite{tri}.}.
Gaugino masses turn to be zero 
at tree level in these theories.
This is because the super-Weyl-K\"ahler
symmetry 
does not allowed for
a coupling between $\varphi$ and the vector superfield which
 transforms as a singlet under this symmetry.
Nevertheless, as we will see in the next section,  gaugino masses can
arise at the one-loop level.

Let us finally comment on 
the supersymmetric mass of the Higgs, $\int d^2\theta \mu H_uH_d$. 
If this term is not allowed in the superpotential (by an $R$-symmetry),
but it is 
present in the K\"ahler,  $\lambda H_uH_d\in K$, 
the first term of 
eq.~(\ref{supergravity}) will generate the coupling
\begin{equation}
\int d^4\theta \varphi\varphi^\dagger (\lambda H_uH_d + {\rm h.c.})\, .
\label{mutermi}
\end{equation}
After supersymmetry breaking, we obtain by replacing 
 eq.~(\ref{ftermphi}) in eq.~(\ref{mutermi})
\begin{equation}
\int d^2\theta\, \mu H_uH_d\, ,
\label{muterm}
\end{equation}
where $\mu =\lambda m_{{3/2}}$ is of the right order of magnitude.
This is just the mechanism of Ref.~\cite{mu}.
Eq.~(\ref{mutermi}) also generates a bilinear soft term for the Higgs, 
$B\mu=-\lambda m^2_{3/2}$.

\subsection{Low-energy soft masses}

Although the  
gaugino masses are zero at $M_P$, they can be generated
at the 
one-loop level
if there are  heavy chiral superfields in the theory \cite{onel}. 
Each heavy superfield gives
a contribution to the gaugino masses at the one-loop level:
\begin{equation}
  \label{oneloop}
  m_{\lambda_a}=-\frac{\alpha_a}{4\pi}\sum_RS_a(R)\, 
m_{3/2}\, ,
\end{equation}
where $S_a(R)$ is the Dynkin index of a heavy superfield
in the $R$  representation 
of the gauge group $G_a$ ($S_a=1/2$ for the fundamental representation).
In GUTs
 the one-loop contribution of eq.~(\ref{oneloop}) can be 
quite large due to the large number of heavy fields.
One   can expect
$m_{\lambda_a}\sim m_{3/2}$. 
In GUTs there are also contributions from the heavy vector multiplets
 \cite{onel}.
These contributions 
do not respect the GUT symmetry, and therefore
provide deviations from
gaugino mass unification.

In the superfield formalism of eq.~(\ref{supergravity}),
the gaugino masses of
eq.~(\ref{oneloop}) 
can  be understood as arising from  
the heavy-field contribution to the 
super-Weyl-K\"ahler anomaly \cite{anomaly}:
\begin{equation}
  \label{anomaly}
\frac{\sum_RS_a(R)}{32\pi^2}\int d^2\theta \ln\varphi W_aW_a\, ,
\end{equation}
where $W_a$ is the superfield strength of the gauge vectors.
Inserting 
 eq.~(\ref{ftermphi}) in the above equation,  we recover
 eq.~(\ref{oneloop}).
Eq.~(\ref{oneloop}) holds at the scale at which the heavy fields
are integrated out;
to obtain the values of the gaugino masses at a low-energy scale, 
one still has to evolve them down using the RGE of the 
supersymmetric standard model.  

The scalar 
 soft masses also  receive radiative corrections 
from the low-energy spectrum.
The degeneracy of  the first and second family squarks  will
not be altered if their Yukawa couplings
remain small below $M_P$.
In theories with  no extra U(1), the scalar masses at the weak scale
will be given by
(neglecting the  contribution coming from the SU(2)$_L\times$ U(1)$_Y$
$D$-term and Yukawa couplings\footnote{For the scalar top
and Higgs, the
effect of the Yukawa couplings should be included. A complete
analysis of these masses and the electroweak breaking
is beyond the scope of this letter.})
\begin{equation}
m^2_{i}=-2m^2_{3/2}
+a_{i}m^2_\lambda(M_G)\, ,
\label{fsg}
\end{equation}
where $a_{i}\sim 5-7$ for the squarks,  $\sim 0.5$ for the
left-handed sleptons
and $\sim 0.15$ for  the  right-handed sleptons;
  $m_\lambda(M_G)$ is the one-loop gaugino mass induced
by the heavy fields that we  assumed to be at
the GUT scale.
The requirement of non-tachionic
 selectrons, {\it i.e.}, $m^2_{\tilde e_{L,R}}>0$,
puts a  strong constraint on the gaugino mass,  $m_\lambda(M_G)\gappeq
 3.6m_{3/2}$.
Since the gaugino mass arises at one-loop, 
this constraint is difficult to satisfy.
Extra contributions to the scalar masses can arise if extra U(1)
are present in the theory as we will consider in the next section.

\subsection{Models with an extra U(1)}
\label{section}

A very well motivated candidate for  a gauge U(1)-symmetry
is the anomalous U(1)
often present in  string theories. The 
anomaly is cancelled 
by the Green-Schwartz mechanism which requires non-zero (and equal)
mixed anomalies for each gauge group \cite{gs}. 
A non-zero gravitational anomaly
results into the appearance
of a Fayet-Iliopoulos (FI)
term  $\xi={\cal O}(M^2_P{\rm Tr}{\bf Q})$.
In string theories the  FI-term 
can be calculated and is given by \cite{fayetil}
$
 \xi = g^2{\rm Tr}{\bf Q}M^2_P/( 192\pi^2)\, .
$
We will assume that Tr${\bf Q}<0$.
Below $M_P$, the potential is given by
\begin{equation}
\label{potentialdterm}
V=-2m^2_{3/2}\sum_i|\phi_i|^2+
\left|\frac{\partial W}{\partial \phi_i}\right|^2+
\frac{g^2}{2}\left(\sum_iq_i|\phi_i|^2+\xi\right)^2+{\cal O}(m_{3/2}^4)\, .
\end{equation}
{}From the minimization of (\ref{potentialdterm}),  
one can see that, among the fields with  $F$-flat directions,
the one with the 
smallest (positive) charge $q$ will get a VEV
\begin{equation}
  \label{vev}
\langle|\phi|^2\rangle=\frac{1}{q}\left(\frac{2m^2_{3/2}}{g^2q}
-\xi\right)\, ,  
\end{equation}
and a non-zero $D$-term will be generated 
 \footnote{An $F$-term
for $\phi$ is also generated in this model,
$F_\phi=m_{3/2}\langle\phi\rangle$,
 but its contribution to the soft masses is 
 small  since $\langle\phi\rangle\ll M_P$.},
$\langle D\rangle=2m^2_{3/2}/(qg)$.
The non-zero $D$-term contributes to the soft masses of all the charged
scalars \cite{scalar,us}
\footnote{ 
This contribution can overwhelm the gravity-induced contribution
if the anomalous U(1) is responsible for supersymmetry breaking 
\cite{us}. This is an alternative way to suppress flavor-violating
interactions.}
 and we have at the scale $\sqrt{\xi}$ 
\begin{equation}
m_i^2 = 2\left(\frac{q_i}{q}-1\right)m^2_{{3/2}}\, .
\label{dtermcont}
\end{equation}
Thus,   fields with
$q_i>q$  will have positive scalar soft masses.
We will require that all matter fields as well as any
field corresponding to
a (exactly) flat direction satisfy 
this requirement and, therefore,  there are not tachionic states 
(alternative scenarios could also be possible).
The simplest model along these lines has the following content:
(1) The superfield $\phi$.
(2) The quark and lepton superfields with universal 
U(1)-charge\footnote{We could also consider a 
flavor-dependent charge assignment. The only crucial
condition, to be consistent with flavor-violating constraints,
is to equally charge the first and second family of
down-quarks.}
$q_Q$ larger than  $q$. 
(3) Two Higgs doublets $H_u$ and $H_d$;  the charge of $H_u$
is taken to be  $-2q_Q$ to allow for
a top-quark mass being generated from a renormalizable interaction
$H_uQ\bar U$ (the charge of $H_d$ is more model dependent).
(4) Extra (heavy) superfields 
$\Psi$ and $\bar \Psi$ vector-like under the standard model gauge group. 
The motivation for these  states is two-fold. 
{}First, they are required 
for the mixed anomalies, and secondly they will 
contribute to the gaugino masses in accordance with the 
discussion in the 
previous section. 
They  will get masses from their coupling to $\phi$:
\begin{equation}
\frac{ \phi^n}{ M_P^{n-1}} {\Psi\bar \Psi}\ ,
\ \ \ n=-\frac{q_{\Psi}+q_{\bar\Psi}}
{q}\, .
\end{equation}

In these models the $\mu$-parameter can also be generated 
from a higher dimensional
coupling of $H_uH_d$ to $\phi$.
{}For example, if $H_u$ and $H_d$ 
carry equal charges (in this case
the bottom mass is also generated from a renormalizable
 interaction), the supersymmetric term 
\begin{equation}
\frac{\phi^n}{M_P^{n-1}}H_uH_d\ ,\ \ \
n=\frac{4q_Q}{q}\, ,
\label{mutermb}
\end{equation}
 gives rise to a realistic value of
  $\mu$ 
for $\sqrt{\xi}/M_P \sim 0.1 - 0.01$ and 
$q_Q \sim (2$--$3)q$. 
Alternatively, the 
$\mu$-term could
 be generated from a coupling of $H_uH_d$ to
a neutral singlet $S$ \cite{singletcase,onel}:
\begin{equation}
W = SH_uH_d + S^3\, .
\label{singlet}
\end{equation}
(The cubic term is necessary for eliminating an unwanted 
weak-scale axion).
Since $S$ is neutral under
the anomalous U(1), its soft mass 
is negative.   
This drives a  VEV for $S$ that
generates the  $\mu$-parameter.

Similar scenario could arise with non-anomalous U(1).
An anomaly-free U(1)
is easy to obtain,
since the $\Psi$-$\bar \Psi$ states and matter states
give  opposite contributions to the anomalies.
In this case, a  FI-term of order $\xi\sim m_{3/2}M_P$ 
can be generated
at the one-loop level if there is a 
mixing between  the U(1) $D$-term
and the $D$-term that cancels  the cosmological constant.
This occurs whenever 
Tr${\bf QQ^{'}}\not=0$ where ${\bf Q}$ and 
${\bf Q^{'}}$ are the generators associated to the two $D$-terms.

\section{Phenomenological Implications}

The most important implication of the models with 
vanishing $F$-terms is that they lead to a  mass degeneracy between
 the first and second family of squarks,
and avoid dangerous flavor-violating interactions.
It is also important to remark that 
these scenarios  allow for gauging flavor symmetries $G_f$
without inducing a VEV for the
 $D$-term of $G_f$ (these $D$-terms are very dangerous since
 they usually break
 the interfamily degeneracy). 
A non-zero $D$-term is avoided if
 the fields that break $G_f$ have
universal soft masses; 
this can be the case in our scenarios.

Another
implication of the models
is that the soft terms depend only on one parameter, the gravitino mass 
$m_{3/2}$. 
This allows for a solution to the supersymmetric
CP problem \cite{edm}.
The  origin of this problem resides
in 
the phases of 
the soft terms
and
$\mu$-parameter that 
  induce, at the one-loop level,  too large
electric dipole moments.
Even  the minimal supergravity model suffers from this problem.
In our models,  however, 
all the phases can be rotated away.
If $\mu$ arises from a supersymmetric term,
like in eq.~(\ref{muterm}), (\ref{mutermb}) or (\ref{singlet}),
its phase
can be eliminated by a rotation of the Higgs superfields.
In the soft terms the only possible phase appears in
$m_{3/2}$ that can be eliminated by an $R$-rotation.

The model of section~\ref{section} also allows 
for a solution to the strong CP problem.
If $\mu$ is induced from eq.~(\ref{mutermb}), 
there is a remnant Peccei-Quinn symmetry
broken by the VEV of $\phi$.
The axion is the superposition 
of  the imaginary part of $\phi$  and the dilaton,
orthogonal to the one
eaten up by the gauge U(1) field.

A final  important feature of 
models with vanishing $F$-terms is that they do not suffer
from the Polonyi
problem \cite{polonyi}.
This problem arises in conventional supergravity theories
because supersymmetry  is broken by non-vanishing $F$-terms of the hidden
sector fields.
In our case, $F$-terms  vanish and the Polonyi problem does not
arise (the only field  that gets a non-zero
$F$-term is $\varphi$ that is not a dynamical field). 

\section{Conclusions}

We have considered a class of supergravity theories in which the
hidden superfields have vanishing $F$-terms but 
a non-zero superpotential. 
This is  generated dynamically at a low-energy  and
 is the source of
supersymmetry breaking in the observable sector
as soon as the
 cosmological constant is set to zero by a non-zero $D$-term.

The gravity-induced  soft supersymmetry breaking terms 
depend only on the gravitino mass 
and not on the details of the hidden sector.
We find 
\begin{eqnarray}
&&m^2_{i\bar\jmath}=-2m^2_{3/2}\delta_{i\bar\jmath}\, ,\nonumber\\  
&&A_{ijk}=0\, ,\nonumber\\  
&&B_{ij}=-m_{3/2}M_{ij}\, ,\label{final}
\end{eqnarray}
for the scalar masses, trilinear and
bilinear soft terms respectively.
This universal pattern of soft masses guarantees the absence of
flavor-violating interactions. No 
 new CP-violating phases appear.
The mechanism of  ref.~\cite{mu} to generate a 
 $\mu$-parameter 
is also operative in these theories. 

The negative soft masses in eq.~(\ref{final})
can become positive at low-energies
if extra U(1) are present in the theory. We have considered 
models in which the matter fields transform under an anomalous U(1).
This is often the case in string  theories. In such cases, the soft masses
receive a positive contribution (see eq.~(\ref{dtermcont})).
The gaugino masses are zero at $M_P$, but they are induced 
at the one-loop level 
\begin{equation}
  m_{\lambda_a}=-\frac{\alpha_a}{4\pi}\sum_RS_a(R)\, 
m_{3/2}\, ,
\end{equation}
if heavy states 
between $M_P$ and the weak scale,
transforming under the standard model group,
are present in the theory.
Thus, 
 the gaugino masses   carry us information about
the GUT physics.

Unlike conventional supergravity models, there are not
cosmological problems in the hidden sector  since there are not 
Polonyi-like states with large VEVs and small masses.

\vspace{1cm}

It is a pleasure to thank R. Barbieri, E. Dudas, R. Rattazzi 
and C. Savoy  for very useful discussions.


\newpage

\begin{thebibliography}{99}

\bibitem{sugra}
R. Barbieri, S. Ferrara and C.A. Savoy, Phys. Lett. {\bf B119} 
(1982) 343;
A.H. Chamseddine, R. Arnowitt and P. Nath, 
Phys. Rev. {\bf D49} (1982) 970.

\bibitem{soni} S.K. Soni and H.A. Weldon, 
Phys. Lett. {\bf B119} (1982) 343.

\bibitem{edm} W. Buchm\"uller and D. Wyler, 
Phys. Lett. {\bf B121} (1983) 321;
J. Polchinski and M.B. Wise, Phys. Lett. {\bf B125} (1983) 393.

\bibitem{sugrapo}
E. Cremmer {\it et al.},  Phys. Lett. {\bf B79}
(1978) 231;  Nucl. Phys. {\bf B147} (1979) 105; E. Witten and J.
Bagger, Phys. Lett. {\bf B115} (1982) 202; E. Cremmer,S. Ferrara,
L. Girardello and A. van Proeyen, Phys. Lett. {\bf B116} (1982)
231; Nucl. Phys. {\bf B212} (1983) 413; J. Bagger, Nucl.
Phys. {\bf B211} (1983) 302.

\bibitem{nonpert}
  For a review see, K. Intriligator and N. Seiberg, hep-th/9509066,
and references therein.

\bibitem{ads} I. Affleck, M. Dine and N. Seiberg, Nucl. Phys.
{\bf B256} (1985) 557;
A.I. Vainshtein, V.I. Zakharov and M.A. Shifman,
 Sov. Phys. Usp. {\bf 28} (1985) 709;
D. Amati, K. Konishi, Y. Meurice, G. Rossi and
G. Veneziano Phys. Rep. {\bf 163} (1988) 169.

\bibitem{savas} 
M. Shifman, hep-th/9704114; 
S. Dimopoulos, G. Dvali and R. Rattazzi, hep-ph/9705348.

\bibitem{string}
L. Iba\~nez and D. Lust, Nucl. Phys. {\bf B382} (1992) 305;
V. Kaplunovsky and J. Louis, Phys. Lett. {\bf B306} (1993) 269;
A. Brignole, L. Iba\~nez and C. Mu\~noz, 
Nucl. Phys. {\bf B422} (1994) 125.

\bibitem{compensator} T. Kugo and S. Uehara, Nucl. Phys. 
{\bf B222} (1983) 125. 

\bibitem{anomaly}
V. Kaplunovsky and J. Louis, Nucl. Phys. {\bf B422} (1994) 57.

\bibitem{bagger}
J. Bagger, E. Poppitz and L. Randall, Nucl. Phys. {\bf B455} (1995) 59.

\bibitem{tri} See, for example,
A. Pomarol and D. Tommasini, Nucl. Phys. {\bf B466} (1996) 3;
R. Barbieri, G. Dvali and L.J. Hall, 
Phys. Lett. {\bf B377} (1996) 76.

\bibitem{mu} G.F. Giudice and A. Masiero, Phys. Lett. {\bf B206} 
(1988) 480.

\bibitem{onel}
 J.P. Derendinger and  C.A. Savoy, Nucl. Phys. {\bf B237} (1984) 307.

\bibitem{gs} M. Green and J. Schwarz, Phys. Lett. {\bf B149} (1984) 117.

\bibitem{fayetil} 
J. Atick, L. Dixon and A. Sen, Nucl. Phys. {\bf B292} (1987) 109;
M. Dine, I. Ichinose and N. Seiberg, Nucl. Phys. {\bf B293}
(1987) 253.

\bibitem{scalar} H. Nakano, hep-th/9404033;
E. Dudas, S. Pokorski and C.A. Savoy,
 Phys. Lett. {\bf B369} (1996) 255;
Y. Kawamura and  T. Kobayashi, Phys. Lett. {\bf B375} (1996) 141;
E. Dudas, C. Grojean, S. Pokorski and
 C.A. Savoy,  Nucl. Phys. {\bf B481} (1996) 85.

\bibitem{us}
G. Dvali and  A. Pomarol, 
Phys. Rev. Lett. {\bf 77} (1996) 3728;
P. Binetruy and
E. Dudas, Phys. Lett. {\bf B389}  (1996) 503.

\bibitem{singletcase}
H.P. Nilles, M. Srednicki and D. Wyler, Phys. Lett. {\bf B120} (1983) 346.

\bibitem{polonyi}
 G.D. Coughlan, W. Fischler, E.W. Kolb, S. Raby and
G.G. Ross, Phys. Lett. {\bf B131} (1983) 59;
J. Ellis, D.V. Nanopoulos and M. Quir\'os, 
Phys. Lett. {\bf B147} (1986) 176.


\end{thebibliography}
\end{document}